\documentclass[a4paper,aps,pra,showpacs,preprintnumbers]{revtex4}
\usepackage{graphics,graphicx,epsfig}
\usepackage[centertags]{amsmath}
\usepackage{amsfonts}
\usepackage{amssymb}
\usepackage[english]{babel}
\usepackage{graphicx}
\usepackage{amsthm,color}
\usepackage{epstopdf}
\usepackage{euscript}

\begin{document}

\def\BE{\begin{equation}}
\def\EE{\end{equation}}
\def\BEA{\begin{eqnarray}}
\def\EEA{\end{eqnarray}}
\def\BY{\begin{eqnarray}}
\def\EY{\end{eqnarray}}

\def\L{\label}
\def\nn{\nonumber}
\def\ds{\displaystyle}
\def\o{\overline}

\def\({\left (}
\def\){\right )}
\def\[{\left [}
\def\]{\right]}
\def\<{\langle}
\def\>{\rangle}

\def\h{\hat}
\def\hs{\hat{\sigma}}
\def\td{\tilde}

\def\k{\mathbf{k}}
\def\q{\mathbf{q}}
\def\r{\vec{r}}
\def\ro{\vec{\rho}}
\def\a{\hat{a}}
\def\b{\hat{b}}
\def\c{\hat{c}}
\def\h{\hat}

\title{Noiseless signal shaping and cluster state generation with quantum memory cell} \vspace{1cm}
\author{A.~D.~Manukhova, K.~S.~Tikhonov, T.~Yu.~Golubeva, Yu.~M.~Golubev}
\address{St.Petersburg State University, 7/9 Universitetskaya nab., St. Petersburg, 199034 Russia}
\date{\today}

\begin{abstract}
In this article, we employ multimode radiation of a synchronously pumped optical parametric oscillator (SPOPO) to build a cluster state through a
conversion on the base of quantum memory cell. We demonstrate that by choosing an appropriate driving field we can ensure the effective writing of
the only one supermode from the entire set of the SPOPO squeezed supermodes. Further, by changing the driving field profile at the readout, we
convert the time profile of the retrieved signal while maintaining its quantum state. We demonstrate the possibilities of using the presented scheme
by the example of creating a four-mode linear cluster state of light.
\end{abstract}
\maketitle

\section{Introduction\L{I}}

The quantum memory cells, being a key tool of quantum communications over long distances, have been studied in detail in various versions of
light-matter interaction \cite{Hammerer, LvovskyRev, Simon, NunnRev}. However, recently increasing attention of researchers is being paid to the use
of memory cells not only as a delay mechanism for a quantum signal, but also for converting this signal and manipulating it directly within the
memory cell \cite{Campbell, Moiseev16}. Viewed in this light, quantum memory becomes an element of a quantum computer circuit.

In this regard, an important aspect of the memory cell's work is its ability to store signals of various shapes. Changing the profile of the driving
field allows one to store effectively signals with different time profiles that is actively used to optimize the memory \cite{Gorshkov2007,
Novikova1, Novikova2}. To increase the efficiency of signal storage it is typically sufficient to optimize the full memory cycle as a whole using the
same driving field during both the writing and readout stages \cite{NunnTheses, Tikhonov2015}. A different situation arises when we want not only to
restore the input signal, but to modify it properly during the storage.

The problem of converting the shape of a quantum signal in the resonator quantum memory scheme was recently discussed in \cite{Sokolov2015, Bimbard,
Stanojevic, Kalachev, Dilley}. Exciting the medium in a way that the signal entered the resonator as efficiently as possible, the authors modified
the driving field  to obtain a signal of a given shape at the output of the resonator. In that case, the specificity of operation in the resonator
configuration made it possible to exclude from consideration the longitudinal spatial aspect of the excitation distribution in the medium. However,
the only slow signals (in comparison with spectral width of the cavity mode) can be modified by this way; in particular, this transformation is
impossible for SPOPO light. We will show that such a conversion of a signal shape can be implemented not only in the resonator model of memory, but
also in a free space. This is the first objective of this work.

As a quantum signal to preserve and convert within the memory cell, we will consider the essentially multimode radiation from an optical parametric
oscillator synchronously pumped by a femtosecond laser (SPOPO) \cite{Patera} in a subthreshold resonator configuration. It is known that this
radiation possesses interesting quantum properties, for instance, the genuine multipartite entanglement has been experimentally demonstrated
\cite{Gerke}. A proper basis of the modes of this light, named "supermodes"\; by the authors (we will follow the same terminology), has been found,
and the presence of six squeezed supermodes  was demonstrated experimentally \cite{Roslund}. Moreover, limiting the number of squeezed supermodes, as
shown in \cite{Roslund, Araujo}, is related to the experimental observation technique, and not to the nature of the radiation. The theory predicts
the presence of about a hundred squeezed independent degrees of freedom of such a light \cite{Patera}. All this makes the SPOPO radiation attractive
from the perspective of creation of multipartite quantum states for quantum computations.

The second objective of this work is to construct a quantum cluster state on the basis of SPOPO radiation. It is well known that cluster states
provide a resource for one-way computations \cite{Raussendorf, Briegel}. The combination of cluster generation mechanism and the quantum memory
protocol allows one to extend the time frames of manipulation in the one-way computational schemes and to overcome the problem of the Gaussian state
decoherence \cite{Campbell, Korolkova, Oxana}. At the same time one should take into account that the preparation of a cluster state assumes mixing
of the light modes and distinguishing the quadrature components in a homodyne process. Without introducing additional noises, such operations are
possible only with the same modes. Therefore, constructing of a cluster with orthogonal modes \cite{Araujo} requires clarification and concretization
of the procedure. To avoid this complexity, we will use the mode conversion procedure, considered in the first part of this article.

\section{Quantum memory scheme for SPOPO radiation}

Before discussing the signal profile conversion in a memory cell, we briefly discuss a quantum memory mechanism and features of the writing and
readout of the pulsed SPOPO radiation.

\subsection{Model}

In this section, a brief description of the quantum memory model based on the off-resonant Raman interaction of the quantum signal and classical
driving fields with an ensemble of three-level atoms in the  $\Lambda$-configuration is presented. Detailed construction of this model, developed by
the same team of authors, can been found in \cite{Manukhova}.

We solve the one-dimensional spatial problem, assuming that all atoms of the ensemble are located within a plane infinite layer of a length $L$
perpendicular to the $z$-axis, along which periodic trains of $N$ pulses of the signal and driving light fields propagate simultaneously during the
writing. The duration of a single pulse in each train is $T_0$, and the period between pulses is $T$. According to the experiment \cite{Pinel}, the
relation $T_0<<L/c<< T$ is met, so that at any given time  not more than one pulse propagates through the medium. During the readout, the driving
runs in the opposite direction, since we consider the case of a backward readout that is much more efficient than a forward one \cite{GorshkovII,
Surmacz, Golubeva2011, Moiseev}.

In the spectral representation, the signal field generated by SPOPO in the process of a parametric light scattering \cite{Patera} has a form of a
frequency comb with the width $\omega_c$ and the distances between the teeth $\omega_{rep}=2\pi/T$, corresponding to the frequency of the ring cavity
round-trip of a correlated photon pair. The spectral structure of the signal will be discussed in more detail in Sec. \ref{source}.

\begin{figure}[h]
 \begin{center}
  \includegraphics[height=45mm]{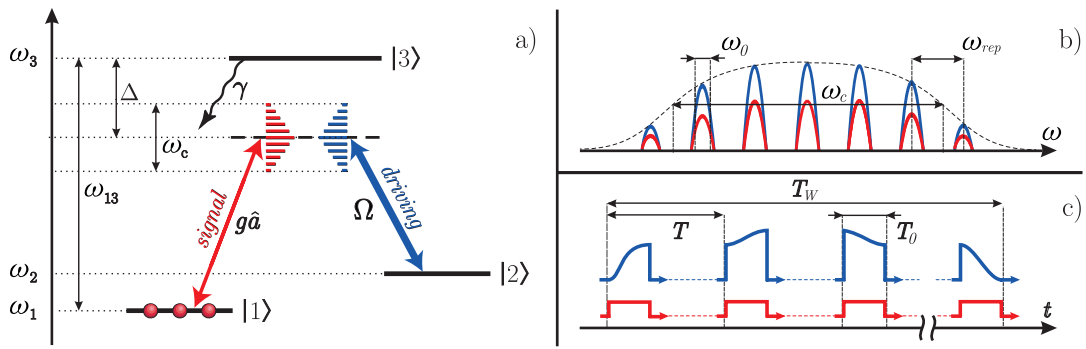}
\caption{a) Energy level scheme of an atomic ensemble, interacting with combs of signal and driving in the Raman memory protocol; b) frequency
profile of the signal and driving fields:   $\omega_c$ -- spectral width of the combs, $\omega_{rep}$ -- distances between the comb's teeth,
$\omega_0$ --  width of a single tooth; c) time profile of the signal and driving fields: $T_W$ --  total duration of the periodic pulse trains, $T$
-- repetition period, $T_0$  --  duration of a single pulse.}
  \label{Raman1}
 \end{center}
 \end{figure}

Figure \ref{Raman1} provides an energy level scheme of the atomic ensemble and the characteristic time and frequency parameters of the light fields
acting on the corresponding transitions.
We assume that initially all the atoms of the ensemble are prepared in the ground states $|1 \rangle$ by optical pumping. Field polarizations are
adjusted so that the signal field acts on the transition between the levels  $|1  \rangle$ and $|3  \rangle$, and the driving field acts on the $|2
\rangle$ - $|3  \rangle$  transition. Carrier frequencies $\omega_s$ and  $\omega_d$ of signal  and driving are detuned from the resonance
frequencies $\omega_{13}$ and $\omega_{23}$ of the atomic transitions, respectively ($\Delta=\omega_s-\omega_{13}=\omega_d-\omega_{23}$). In this
model, we assume $\omega_c \ll |\Delta|$ that allows us to consider the detuning for each tooth of the comb to be the same.

The interaction Hamiltonian in the rotating wave and dipole approximations can be derived in form:

\BY
\hat{V}=&&\int\limits_0^L d z \;  \left( i\hbar g\hat{a}(t,z) \hat{\sigma}_{31}(t,z)e^{-i\Delta t+ik_sz}+ i\hbar\Omega(t,z)\hat{\sigma}_{32}(t,z)
e^{-i\Delta t+ik_dz} +H.c. \right) .\L{1}
\EY
Here, $g=\sqrt{\omega_s/2\varepsilon_0 c \hbar}\;d_{13}$  is the coupling constant between a single atom and the signal field, $k_s$ and $k_d$ are
the wave numbers of the signal and driving, respectively. For simplicity the Rabi frequency $\Omega(t,z)$, through which we describe the driving
field, and the matrix element $d_{31}$ of the dipole moment operator are assumed to be real.

The slowly varying amplitude of the signal field is given by the annihilation bosonic operator $\hat{a}(t,z)$ which obeys the following commutation
relations:
\BY
&& [\hat{a}(t,z),\hat{a}^\dag(t,z')]=c\left (1-\frac{i}{k_s}\frac{\partial}{\partial z}\right)\delta(z-z'),\qquad  [\hat{a}(t,z),\hat{a}^\dag
(t',z)]=\delta(t-t').\L{2}
\EY
At the same time, collective coherence operators $\hat{\sigma}_{31}(t,z)$ and $\hat{\sigma}_{32}(t,z)$ and their conjugates obey the following:
\BY
[\hat{\sigma}_{ij}(t,z),\hat{\sigma}_{ji}(t,z')]=\left[ \hat{\sigma}_{ii}(t,z)-\hat{\sigma}_{jj}(t,z)\right]\delta(z-z'),\L{3}
\EY
where $\hat{\sigma}_{ii}(t,z)$  is the collective operator of the population of level $| i \rangle$.

In \cite{Manukhova}, the simplest case when the time profile of the driving had a form of a periodic sequence of $N$ rectangular pulses with duration
$T_0$ was analyzed. Here, we do not limit ourselves to a specific form of the driving while constructing equations and solutions to be able vary this
parameter further. Taking into account the phase delay  $z/c$  for the driving field, we derive the Rabi frequency as:
\BY
&&\Omega(t,z)=\Omega_0f(t-z/c),\qquad f(t)=\sum_{n=1}^N F(t)\Theta(t-(n-1)T),\qquad  \Theta(t)=H(t)\cdot H(T_0-t). \L{4}
\EY
Here, $F(t)$ is the envelope of the driving pulse train that we will specify further, and $H(t)$ is the Heaviside function. Accordingly, when fields
are switched on simultaneously, each pulse of the driving field corresponds to the appropriate signal one.

The ability to specify the profile of the driving will allow us not only to ensure the best preservation and restoring of quantum statistical
characteristics of the signal field, but also to change the profile of the output signal, that will be shown in Sec. \ref{Change}.

\subsection{The Heisenberg-Langevin equations}

Taking into account the commutation relations (\ref{2}) and (\ref{3})  we can easily obtain the system of Heisenberg equations for the slowly varying
amplitudes. According to the Wigner-Weisskopf theory, the resulting system should be supplemented by relaxation terms and corresponding Langevin
noise sources, which describe the spontaneous decay from the level  $ | 3 \rangle $. Then the system is averaged over the atomic positions. Assuming
that the change of population of the level $ | 1 \rangle $  during the all memory process is negligible and eliminating adiabatically the level
$|3\rangle$, within the Raman limit $|\Delta|\gg\gamma$ we obtain a closed set of two equations \cite{GorshkovII, Polzik}:
\BY
&&\left(\frac{1}{c}\frac{\partial}{\partial t}+\frac{\partial}{\partial z}+i\frac{g^2N_{at}/L}{\Delta}\right)\hat{a}(t,z)=-i\frac{  g\sqrt {N_{at}/L}\;\Omega_0}{\Delta}f(t-z/c)\hat{B}(t,z)+\hat{n}_a(t,z),\L{5} \\
&&\left(\frac{\partial}{\partial t}+i\frac{\Omega_0^2}{\Delta}f(t-z/c)\right)\hat{B}(t,z) = -i\frac{g^2N_{at}/L}{\Delta}f(t-z/c)\hat{a}(t,z)+\hat{n}_b(t,z),\L{6}
\EY
where $\hat{B}(t,z)=\hat{\sigma}_{12}(t,z)/\sqrt{N_{at}/L}$ is the spin coherence operator renormalized to satisfy bosonic commutation relations,
${N_{at}/L}$  is the linear atomic density of the memory cell, $\hat{n}_a(t,z)$ and $\hat{n}_b(t,z)$  are the Langevin noise sources.

Since we solve the problem in time representation, we inevitably consider all nonresonant terms within the width of the comb along with an exact
two-frequency resonance. Such an approach distinguishes this consideration from that of \cite{Oxana}.

\subsection{General solutions for the writing and readout}\L{solutions}

For further convenience, all formulas and expressions will be given in dimensionless variables $z$ and $t$, so the spatial variable is expressed in
units of the optical depth, scaled to $\gamma/\Delta$ , and the time variable  -- in units of the Rabi frequency, scaled to $\Omega_0/\Delta$:
\BY
&&\frac{\Omega_0^2}{\Delta}t\rightarrow t \:,\qquad \frac{g^2(N_{at}/L)}{\Delta}z\rightarrow z  \;.\L{dimless}
\EY

Equations  (\ref{5})--(\ref{6}) describe the evolution of the field and material variables during the times of light-atom interaction. Depending on
the chosen initial conditions, these interactions may correspond to the writing or the readout of the signal.

The writing process of the quantum state of light on the atomic ensemble suggests that the signal field $\hat{a}_{in}(t)$ enters the front face of
the memory cell ($z=0$), and herewith the spin oscillator of the atomic subsystem is initially in the vacuum state. With these initial and boundary
conditions, we solve the set (\ref{5})--(\ref{6}), using Laplace transform method, and obtain the expression for the spin coherence of the atomic
ensemble when the whole pulse train has passed through the medium:
\BY
&&\hat {B}(z)=\int\limits_{0}^{T_W} dt f_W(t)\;\hat a_{in}(t)\,J_0\left(2\sqrt{q_W(t)z}\right) \;+vac,\L{8}
\EY
where
\BY
&& q_W(t)=\int\limits_{0}^{t}dt^{\prime}f_W^2(t^{\prime}), \qquad T_W^{-1}q_W(T_W)=1.\L{9}
\EY
Hereafter, with $vac$ we mean the contributions from the vacuum channels; $T_W=(N-1)T+T_0$ is the writing time of the whole pulse train. Function
$f_W(t)$ corresponds to the time profile of the driving during the writing.

To describe the backward readout, we solve Eqs. (\ref{5})--(\ref{6}) with the other initial and boundary conditions: the quantum mode of the signal
field is in a vacuum state, and the spin wave distribution coincides with that obtained at the end of the writing. Note that we assume the storage
process to be ideal, so the spin coherence during the storage time remains the same \cite{Tikhonov2015}. The expression for the amplitude of the
output signal is as follows:
\BY
&&\hat a_{out}(t)= f_R(t)\int\limits_0^L dz \hat { B}(z)\,J_0\left(2\sqrt{q_R(t)z}\right)\; +vac.\L{10}
\EY
The function $q_R(t)$ corresponds to the time profile of the driving field $ f_R (t) $ during the readout and is defined similar to Eq. (\ref{9}).
For simplicity, we assume that the readout time $T_R$ coincides with the writing time $T_W$.

It should be noted that the exact solutions (\ref{5})--(\ref{6})  generate a phase rotation of the field oscillators. However, it can be quite easily
compensated experimentally by adding phase-shift devices before and after the memory cell, as shown in \cite{Manukhova}. Hence,  Eqs.
(\ref{8})--(\ref{10}) do not contain any exponents associated with these phase shifts.

Note that both the writing and the readout processes depend on the field structure determined by the pulse character of the signal and driving
trains. However, in the section \ref{envelopes} we will show that to describe these processes one can use the envelopes of these pulsed fields, since
the quantum correlations of the signal, that are of interest here, appear at times of the order of $T$.

\subsection{Structure of the SPOPO radiation}\L{source}

As a signal field we consider a frequency comb produced in the process of a degenerate parametric light scattering of a periodic train of coherent
pulses with duration of a single one $T_0$, on a $\chi^2$-crystal within a ring cavity \cite{Patera}. The repetition period of the train $T$
($T_0<<T$) coincides with the cavity round-trip time $T$ of the correlated photon pairs of the signal generated in this process. Due to the resonator
Q-factor, photons of the signal can make about ten round-trips over the cavity before leaving, taking part in the parametric generation of the
following photon pairs when they pass through the crystal each next time. This creates correlations in the SPOPO radiation at times of the order of
the several cavity round-trip times $T$.

Generation of the SPOPO’s photons in the process of parametric downconversion is described by the following effective Hamiltonian:
\BY
&&\h{H}=i\hbar \Gamma\sum_{m,n}L_{m,n}\hat a^\dag_m\hat a^\dag_n+H.c. \L{11}
\EY
Here the coupling constant $\Gamma$ is proportional to the pump amplitude and describes the resulting efficiency of the process, $\hat a^\dag_m$ is
the photon creation operator associated with the mode of frequency $\omega_m$.

Matrix $L_{m,n}$ describes the coupling between modes with frequencies $\omega_m$ and $\omega_n$. It is a huge matrix containing about $10^5 \times
10^5$ components. However, by its diagonalization one can reveal eigen degrees of freedom (eigenvectors, or, in other words, Schmidt modes) of the
system, called "supermodes"\ by the authors. Herewith, the analysis, which had been carried out in \cite{Patera}, showed that only a few (about $100$
in theory) eigenvalues, corresponding to the distinguished supermodes, are nonzero. So the double summation in Eq. (\ref{11}) reduces to a single one
for the modes with nonzero eigenvalues only. Supermode profiles in both time and frequency domains can be approximated by Hermite-Gaussian functions
with a high accuracy. Thus, the output radiation of the SPOPO can be represented as a Hermite-Gaussian polynomial expansion:
\BY
&& \hat a_{in}(t)=\sum_k L_k(t)\hat e_k,\L{12}
\EY
where $L_k(t)$ is the Hermite-Gaussian function of order $k$, and the coefficient $\hat e_k$ is the annihilation operator of the photon in the $k-$th
mode, which obeys the standard commutation relations.

The introduced supermodes are found to be very useful under analysis of correlation properties of the SPOPO radiation, not only because of the great
simplification of the mathematical apparatus, but also revealing the genuine quantum degrees of freedom of such a complex system.

Experimentally, the first six supermodes were observed in a squeezed quantum state \cite{Roslund}, therefore, we will follow the possibility of
manipulating precisely these supermodes. Note that further analysis can be carried out for a larger number of supermodes. This fact is essential
since by upgrading the experimental setup (e.g., better adapting the pump spectrum, increasing cavity bandwidths, etc.) the number of observed
squeezed supermodes can be increased up to $\sim 100$ \cite{Araujo}.

\subsection{Passage from a pulsed structure to an envelope}\L{envelopes}

As was shown in the previous section, the characteristic features of the SPOPO radiation can be  well described by smooth Hermite-Gaussian functions
(\ref {12}). At the same time, we do not consider correlations within each single pulse of the radiation, whereas they are very small compared to
correlations between pulses, as it was shown in \cite{Averchenko2011,Averchenko2011a}. Then it is possible to simplify the solutions of the
Heisenberg-Langevin equations (\ref{8}) and (\ref{10}) and consider only the envelopes of the signal and driving instead of their complex pulsed
structure.

Before discussing a mathematical procedure of the passage to the envelopes, it has to be said that our further analysis will be concerned only with
the averages of the normally ordered operators, thus we can omit the contribution from the vacuum channels in Eqs. (\ref{8})  and (\ref{10}).
Besides, the signal and driving fields are time synchronized and each pulse of the driving corresponds to the appropriate signal one. Therefore,
making a simultaneous passage to their envelopes, we should trace with accuracy the preservation of the commutation relations and the normalization.

We will describe the passage from the time profile $f(t)$ with the pulsed structure to its envelope $F(t)$ on the example of the driving field
(\ref{4}). In the case of a signal field, the procedure appears to be similar.

First of all, let us ensure that the normalization condition (\ref{9}) remains satisfied. Figure \ref{Fig2} represents by a blue solid line the
square of the function $f(t)$ for a single pulse.
\begin{figure}[h]
\includegraphics[height=4.0cm]{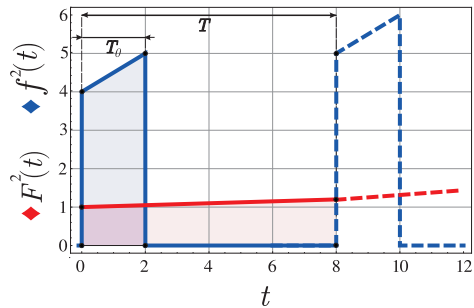}
\caption{ A graphic presentation of the passage from the pulsed structure of the fields to their envelops. }\L{Fig2}
\end{figure}
We assume the number of pulses under the envelope to be large, i.e. $N\gg1$, so $f^2(t)$ under each individual pulse can be approximated by a
straight line. Thus, in order to satisfy the normalization condition after the passage, the height of the trapezium (red solid line) should be
increased in $T/T_0$ times and both of its bases should be decreased accordingly. Hence the envelope function $F(t)$ has to be multiplied by the
normalization factor $\sqrt{T_0/T}$. Due to  the character of the solutions (\ref{8}) and (\ref{10}), the larger base of the trapezium of a pulse
coincides with the smaller base of that of the next one. Extending the procedure for the remaining pulses, we get a polygonal line that can be easily
approximated by a smooth curve. Exactly this curve is the desired normalized envelope of the driving field.

Let us introduce the normalization factor in the definition of functions,
\BY
&& \sqrt{T_0/T} F(t)\rightarrow F(t),\qquad  \sqrt{T_0/T} \hat{A}_{in}(t)\rightarrow \hat{A}_{in}(t),\qquad  \sqrt{T_0/T}\hat{A}_{out}(t)\rightarrow
\hat{A}_{out}(t),\nn
\EY
and derive the solutions (\ref{8}) and (\ref{10}) through the envelopes:
\BY
&&\hat {B}(z)=\int\limits_{0}^{T_W} dt \;\hat A_{in}(t)\,G_{ab}(t,z)
\;+vac, \qquad G_{ab}(t,z)= F_W(t)J_0\left(2\sqrt{Q_W(t)z}\right),\L{15}\\
&&\hat A_{out}(t)= \int\limits_0^L dz \hat { B}(z)\,G_{ba}(t,z)\; +vac, \qquad G_{ba}(t,z)= F_R(t)J_0\left(2\sqrt{Q_R(t)z}\right).\L{16}
\EY
Functions $Q_W(t)$ and $Q_R(t)$  here are determined analogously to Eq. (\ref{9}), but by $F_W(t)$ and $F_R(t)$, correspondingly. Note that their
normalization condition remains the same due to the inclusion of the scale factor $\sqrt{T_0/T}$. Moreover, it is easy to show that they closely
approximate $q_W(t)$ and $q_R(t)$, that means Bessel functions, corresponding to them, also differ little.

The obtained integral transformations (\ref{15}) and (\ref{16})  are written for smooth continuous functions. Their further analysis will be carried
out in Sec. \ref{kernels}, where we will consider the properties of their kernels $G_{ab}(t,z)$ for the writing and $G_{ba}(t,z)$ for the readout.

\subsection{Shaping of the driving for the efficient writing of a single supermode}\L{iteration}

If the writing and the reading are performed with the same driving field $F_W(t)=F_R(t) \equiv F(t)$, it is convenient to introduce a kernel of
integral transformation for the full memory cycle $G(t,t^\prime)$, which links the initial signal field at the input of the cell $\hat
A_{in}(t^\prime)$ with the restored one at its output $\hat A_{out}(t)$:
\BY
&&\hat A_{out}(t)= \int\limits_0^{T_W} dt^\prime \hat A_{in}(t^\prime)\,G(t,t^\prime)\; +vac,\qquad  G(t,t') = \int\limits_0^L dz\;
G_{ab}(t^\prime,z) G_{ba}(t,z). \L{56}
\EY
Since the "half-cycle"\; kernels are equal to each other, the kernel $G(t,t^\prime)$ is symmetrical with respect to permutation of the arguments, so
it is representable as the Schmidt mode expansion \cite{NunnTheses, Fedorov}:
\BY
&& G(t,t') = \sum_{k=1}^{\infty} \sqrt{\lambda_k}\varphi_k(t)\varphi_k(t'), \L{456}
\EY
where $\varphi_k(t)$ is the $k-$th Schmidt mode, and $\sqrt{\lambda_k}$ is the corresponding eigenvalue.

Given that the form of the kernel $G(t,t^\prime)$ depends on the shape of the driving $F(t)$ and the field-medium interaction time $T_W$, we can
change the mode structure of the memory by varying these parameters. Usually, the memory settings are chosen to store as many modes of the signal as
possible, that is, as many eigenvalues as possible in the expansion (\ref{456}) should be close to (or better equal to) one and the rest differ
little from zero \cite{NunnTheses, Manukhova}. Such a multimode operating regime helps to increase the information capacity of the
cell. However, here we seek to implement the "converter"\; for the mode profile of the signal field, so the multimode operating regime does not fit.
We should choose the interaction parameters ($ F (t) $ and $ T_W $) in such a way that only one supermode is effectively written to the memory cell,
and the remaining ones would not interact with it. Mathematically, this amounts to the fact that the series (\ref{456}) has to be represented by only
one term $(\lambda_1=1)$ with the remaining terms zeroed out since the corresponding eigenvalues are equal to zero. Moreover, the profile of the
eigenmode of the cell $\varphi_1(t)$ has to coincide with the profile of the supermode $L_i(t)$ to be written:
\BY
&& G(t,t') = \varphi_1(t)\varphi_1(t')=L_i(t)L_i(t'), \L{iter1}
\EY
We should note that it is not always possible to apply this representation. It is satisfied exactly only if the dimensionless cell length $L$ can be
considered infinitely large compared with the interaction time $T_W$ (expressed in dimensionless units (\ref{dimless})). However, as we shall show
later, even with $L\approx T_W$, one can write and then restore a single chosen mode of the field with a very high accuracy.

To find an appropriate driving field, the following iterative procedure is to be used. Substituting explicitly the kernel (\ref{56})  into the
equality (\ref{iter1}), we put $ t = t^\prime $ and get:
\BY
&& L_i^2(t)= F_i^2(t)\int_0^L dz\;\left[J_0\left(2\sqrt{z\int_0^t dt_1 \;F_i^2(t_1)}\right)\right]^2. \L{iter2}
\EY
Here, the index $i$ points to the number of the supermode, for the writing/readout of which we search the driving. Thus, we can define the driving
profile at the $ j$th step of the iteration $F_i^{(j)}(t)$:
\BY
&&F_i^{(j)}(t) = L_i(t) \left(\int_0^L dz\;\left[J_0\left(2\sqrt{z\int_0^t dt_1 \;(F_i^{(j-1)}(t_1))^2}\right)\right]^2\right)^{-1/2}.\L{iter3}
\EY
$F_i^{(0)}(t)=L_i(t)$ is taken as a zero iteration. Such an iterative procedure converges rapidly, and already at the ninth step the difference
between the obtained and required forms of the kernel $G(t,t^\prime)$ becomes of the order of few percents.

Let us note that the calculation can be substantially simplified if we define the Bessel function by its series expansion
\BY
&& J_0(x)=  \sum_{m=0}^{\infty}\frac{(-1)^m}{(m!)^2}(x/2)^{2m}. \L{iter4}
\EY
Substituting the expansion (\ref{iter4}) into Eq. (\ref{iter3}), we can directly integrate the expression over $z$ and get
\BY
&&F_i^{(j)}(t) =  \sqrt{T_W} L_i(t)\left(\sum_{n=0}^{\infty} \sum_{m=0}^{\infty}\frac{(-1)^{m+n}}{(m!n!)^2} \frac{(L T_W)^{m+n+1}}{m+n+1}
\left(\frac{1}{T_W}\int_0^t dt_1 \;(F_i^{(j-1)}(t_1))^2\right)^{m+n}\right)^{-1/2}.\L{iter5}
\EY
Replacing indexes of summation in Eq. (\ref{iter5}), one of the summations can be performed analytically:
\BY
&& \frac{(L T_W)^{k+1}}{k+1}\sum_{m=0}^k \frac{1}{((k-m)!m!)^2}=\frac{4^k\left(k-\frac{1}{2}\right)!}{\sqrt{\pi}(k!)^3}\frac{(L T_W)^{k+1}}{k+1}
\equiv C_k.\L{iter6}
\EY
Then Eq. (\ref{iter5})  can be derived as follows:
\BY
&& F_i^{(j)}(t) =  \sqrt{T_W} L_i(t)\left( \sum_{k=0}^{\infty}(-1)^{k} C_k A_k(t)  \right)^{-1/2}, \L{iter7}
\EY
where
\BY
&& A_k(t)=  \left(\frac{1}{T_W}\int_0^t dt_1 \;(F_i^{(j-1)}(t_1))^2\right)^{k}.\L{iter8}
\EY
Calculating the values of the coefficients $C_k$ for the chosen parameters $T_W$ and $L$, it is not difficult to find out how many terms of the
alternating series (\ref{iter7}) need to be kept to perform numerical calculation.

\begin{figure}[h]
\begin{center}
\includegraphics[height=40mm]{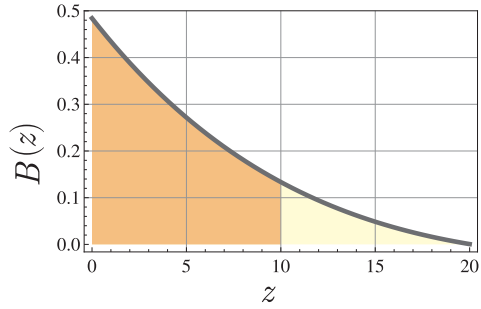}
\caption{Spatial distribution of the spin wave upon excitation of  the medium by SPOPO radiation in the presence of the driving field found by
iterative procedure (\ref{iter7}) before (dark yellow area at $L< 10$)) and  after correction of $L$.} \label{otklik}
\end{center}
\end{figure}

We have presented here an iterative procedure to find a driving field, that ensures the writing and readout of a single chosen supermode of the
signal. As already noted above, in the case of finite values $L$ this procedure is not exact. Nevertheless, let us apply it for calculation at $L=10,
T_W=9$, review to what errors that leads, and show how to reduce them. Carrying out the calculation with the found profile of the driving, one can
verify that, with the parameters discussed, about $9\%$ of the field excitations are not converted to the spin wave of the medium but are lost as a
leakage. One can see from Fig. \ref{otklik} that the coherence at the output face of the cell (for $L=10$) is nonzero.
Because of the multiple rewriting of the field, part of the signal photons leaves the medium: one can say that the length of the medium is
"imperfect"\; for their rewriting. The natural solution is to increase $L$  relative to that one for which the profile of the driving was calculated.
Basing on the shape of the curve in Fig. \ref{otklik}, one can see that it is sufficient to extend the medium to $2L$ so that the coherence at the
output face of the cell becomes zero. More precisely, we do not need to increase the length of the cell, we just have to search for the  desired
driving shape, using  the values  $L$ twice smaller  than  real ones. This ratio is maintained for all considered ranges of $L$ and $T_W$,
corresponding to the experimentally feasible values. Looking ahead, we can say that such a correction of the iterative procedure allows us to
restore/transform the signal amplitude with an accuracy of about $95\%$.

\section{Signal shape conversion on the base of quantum memory cell}\L{Change}

\subsection{Mixing orthogonal modes on a beamsplitter}

Before implementing the idea of a noiseless shaping of the signal profile with the help of quantum memory cell, let us turn once again to the
motivation of this research.

The idea of using orthogonal squeezed supermodes of SPOPO radiation to create a light cluster seems very attractive \cite{Araujo}.

As is well known, a cluster state based on squeezed fields can be constructed by  linear optics transformations, such as mixing fields on
beamsplitters. However, in the case of mixing two orthogonal modes, each of which is squeezed, it is impossible to achieve effective entanglement
through a beamsplitter. Indeed, in contrast to the "traditional" beamsplitter mixing of two squeezed in orthogonal quadratures fields (that leads to
a good entangled state), when the modes are orthogonal to one another, we have to consider the problem of mixing not two but four fields, two
squeezed and two vacuum ones. Such a scheme can generate a state with no more than $50$\%-entanglement.
\\
\\
Let us show it by the example of mixing orthogonal squeezed fields $\hat{S}_1(t)$ and $\hat{S}_2(t)$, wherein the orthogonal modes $L_1(t)$ and
$L_2(t)$ are excited, respectively, and the remaining ones are in the vacuum state:
\BY
&&\hat{S}_1(t)=L_1(t)\hat{e}_1+{vac}, \\
&&\hat{S}_2(t)=L_2(t)\hat{e}_2+{vac}.
\EY
We assume that field $\hat{S}_1(t)$ is squeezed in the amplitude quadrature ( $ \langle  \hat{X}_1^2 \rangle \rightarrow 0$) and the field
$\hat{S}_2(t)$ is squeezed in the phase one ($ \langle \hat{Y}_2^2 \rangle \rightarrow 0$). Thus, a light field with one of the modes being ideally
squeezed, and the remaining ones -- in the vacuum state falls to each input of a beamsplitter. It is easy to follow the field transformations at the
beamsplitter:
\BY
&&\hat{E}_1(t)=\frac{1}{\sqrt{2}}\left[ \hat{S}_1(t)+\hat{S}_2(t)\right],\\
&&\hat{E}_2(t)=\frac{1}{\sqrt{2}}\left[ \hat{S}_1(t)-\hat{S}_2(t)\right].
\EY
Obviously, the squeezed mode in each of the channels is added by a vacuum signal from another one.

Let us estimate the entanglement between oscillators in the mode ${L}_1(t)$  at two outputs of the beamsplitter, using the Duan criterion:
\BY
&&D=\langle (\hat X_1 + vac)^2 \rangle+\langle (\hat Y_1 + vac)^2  \rangle  >\;1/2.
\EY
For the second excited mode the situation is the same. Due to the fact that during the beamsplitter mixing the vacuum noise is added to each of the
modes, the Duan criterion is no longer satisfied, and  each mode of the radiation at the output is half-squeezed and half-entangled. Thus, the
cluster implementation scheme based on orthogonal modes \cite{Araujo} requires additional refinement of the procedure.

As one of the possible ways to overcome this problem, in the next section we will consider the conversion of the mode profile (with preservation of
its quantum state) with the help of the memory cell.

\subsection{Writing and readout of the orthogonal modes}\L{kernels}

Previously, in Eqs. (\ref{15}) and (\ref{16}) we introduced the kernels of integral transformations for the writing $G_{ab}(t,z)$ and the readout
$G_{ba}(t,z)$. Now, taking into account the shapes of the driving $F_i(t)$ that provides the interaction of the memory cell with only one signal mode
with Hermite-Gaussian profile $L_i(t)$ (according to the procedure in Sec. \ref{iteration}), we define two sets of the kernels $G^{(i)}_{ab}(t,z)$
and $G^{(i)}_{ba}(t,z)$:
\BY
&&\hat {B}(z)=\int\limits_{0}^{T_W} dt \;\hat A_{in}(t)\,G^{(i)}_{ab}(t,z)
\;+vac, \L{15_1}\\
&&\hat A_{out}(t)= \int\limits_0^L dz \hat { B}(z)\,G^{(i)}_{ba}(t,z)\; +vac.,\L{16_1}
\EY
where
\BY
&&G^{(i)}_{ab}(t,z)=G^{(i)}_{ba}(t,z)= F_i(t)J_0\left(2\sqrt{z \int\limits_{0}^{t}dt^{\prime}F_i^2(t^{\prime})}\right).\L{16_2}
\EY
The Schmidt decomposition technique can be used not only for the full memory cycle, but also for the half cycles. Let us define the $k-$th response
function of the medium as the spatial profile of the spin wave that appears within the medium when signal field with time profile of the $k-$th
Schmidt mode has been written there:
\BY
&&\sqrt[4]{\lambda^{(i)}_k} g^{(i)}_k(z)=\int\limits_{0}^{T_W} dt \; \varphi^{(i)}_k(t)\,G^{(i)}_{ab}(t,z),\L{457}
\EY
where the index $(i)$ indicates that eigenfunctions and eigenvalues belong to the kernel with the corresponding index.

It is easy to show \cite{Tikhonov2015} that the orthonormality of the eigenfunctions $\{\varphi^{(i)}_k(t)\}$ impose the orthonormality of the
response functions $\{g^{(i)}_k(z)\}$; then the kernels $G^{(i)}_{ab}(t,z)$ and $G^{(i)}_{ba}(t,z)$ can be represented as a series
\BY
&& G^{(i)}_{ab}(t,z) =G^{(i)}_{ba}(t,z)= \sum_{k=1}^{\infty} \sqrt[4]{\lambda^{(i)}_k} g^{(i)}_k(z)\;\varphi^{(i)}_k(t). \L{458}
\EY
Since the driving fields are chosen so that each kernel has only one eigenvalue $\lambda^{(i)}_1$  equal to $1$, and $\lambda^{(i)}_k=0, k\neq 1$,
and the eigenfunctions, corresponding to the nonzero eigenvalue, coincide with the Hermite-Gaussian functions, then  Eq. (\ref{458})  turns to:
\BY
&& G^{(i)}_{ab}(t,z) =G^{(i)}_{ba}(t,z)=  g^{(i)}_1(z)\;L_i(t). \L{459}
\EY
The most interesting property of the transformation under consideration is that we can remove the index "$i$"\;  of the response function on the
right-hand side of Eq. (\ref{459}), since the generated spatial distribution of the spin waves is the same for all driving fields $F_i(t)$, which
have been found by means of the above iterative procedure. This spin wave distribution coincides with $B(z)$ shown in Fig. \ref{otklik}):
\BY
&& G^{(i)}_{ab}(t,z) =G^{(i)}_{ba}(t,z)=  g_1(z)\;L_i(t). \L{460}
\EY
It should be noted that such a situation occurs not for all Hermite-Gaussian functions. Numerical calculation shows that for the chosen parameters
($L=10,T_W=9$) it can be argued with high accuracy that Eq. (\ref{460}) is valid for the first six Hermite-Gaussian modes. By increasing the duration
of the interaction $T_W$ one can increase the number of actual modes, for which Eq. (\ref{460}) is correct and, therefore, for the profiles of which
we can organize a shape converter, as will be shown below. However, one should remember that in real systems we are limited to very specific device
parameters. In particular, elongation of the pulse train in SPOPO leads to a mismatch of the periodic structure of the radiation and, consequently,
to the loss of its quantum properties.

\begin{figure}[h]
\begin{center}
\includegraphics[height=55mm]{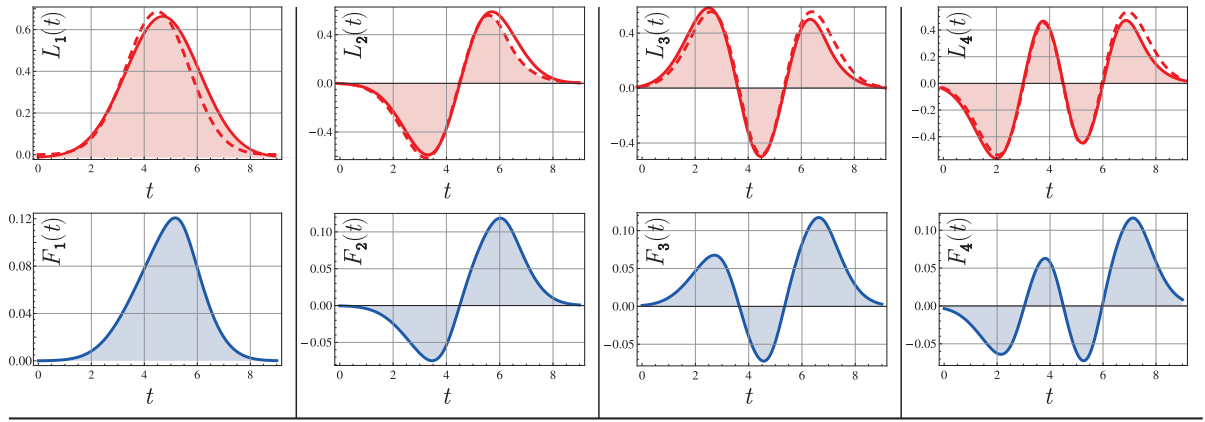}
\caption{The profiles of the first four SPOPO supermodes (upper row, red dashed curves) and the amplitudes $\hat A_{out}(t)$ (upper row, red solid
curves) restored with an appropriate choice of the driving field shapes (bottom row, blue curves) found according to the iterative procedure
\ref{iteration}. } \label{functions}
\end{center}
\end{figure}

Figure \ref{functions} shows the results of a numerical calculation based on the procedure described above.
For each SPOPO supermode, we found the shape of the driving field, that ensured the writing of this particular mode to the memory cell. We made sure
that the response of the medium under such excitation does not depend on the mode number (see Fig. \ref{otklik}) and provided the readout of the
signal with the same driving field (see Fig. \ref{functions}). The results of the calculations show that the accuracy of the signal restoration
(estimated as the overlap integral of the initial and reconstructed modes) is $95.3\%$ for the first four modes.

\begin{figure}[h]
\begin{center}
\includegraphics[height=70mm]{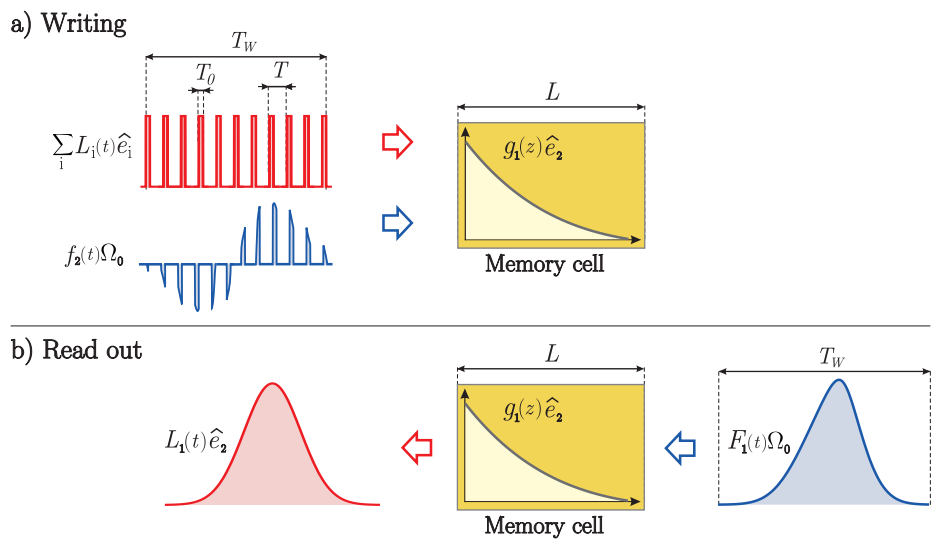}
\caption{Schematic representation  of the shape converter for the signal field.}
\label{Fig4}
\end{center}
\end{figure}

Let us now follow the conversion of the signal shape by the example of writing the second supermode of the SPOPO and the recovering the signal with
the profile of the first Hermite-Gaussian polynomial (see Fig. \ref{Fig4}).
SPOPO radiation  $\hat a_{in}(t)$, constituting the superposition (\ref{12}) of quantum modes with Hermite-Gaussian profiles, enters the memory cell.
According to Eq. (\ref{460}),  in order to write only the second supermode from the entire set of modes, we need to chose the driving field in a form
$F_2(t)$ defined by Eq. (\ref{iter7}). The interaction of these two fields with the atomic ensemble results in the creation of a spin wave with
spatial distribution $g_1(z)$  (see Fig. \ref{Fig4}), whose quantum statistics reproduces the statistics of the scattered supermode $\hat e_2$  of
the signal field. According to Eq. (\ref{460}), during the readout driving field, selected now as a function $F_1(t)$, interacts with the same
excited spin wave, that leads to an effective reading of its quantum-statistical properties to the output signal mode with the profile $L_1(t)$:
\BY
&&\hat A_{out}(t)= \int\limits_0^{T_W} \int\limits_0^L dz dt^\prime \; \hat A_{in}(t^\prime)\, G^{(2)}_{ab}(t^\prime,z) G^{(1)}_{ba}(t,z)\nn\\
&&=\int\limits_0^{T_W} \int\limits_0^L dz dt^\prime \; \sum_k L_k(t^\prime)\hat e_k \; g_1(z)L_2(t^\prime)\; g_1(z)L_1(t)= L_1(t)\hat e_2.
\EY
Thus, we have implemented the conversion of the profile of signal with preservation of its quantum statistics. Numerical calculations show that the
process of signal conversion proceeds with the same accuracy as its restoration discussed above.

\section{Constructing a cluster state based on SPOPO radiation}

In Sec. \ref{iteration} we discussed the possibility of writing each squeezed supermode of the  SPOPO to its memory cell, and in Sec. \ref{kernels}
we considered the transformation of the SPOPO signal profile with preservation of its quantum state. Let us apply the obtained results to construct a
light linear cluster state consisting of four nodes based on the SPOPO light. We dwell upon this example as  the simplest nontrivial cluster state.
It should be noted that the possibilities of constructing a cluster based on the multimode light under consideration are not limited to this state
topology.

\begin{figure}[h]
\begin{center}
\includegraphics[height=100mm]{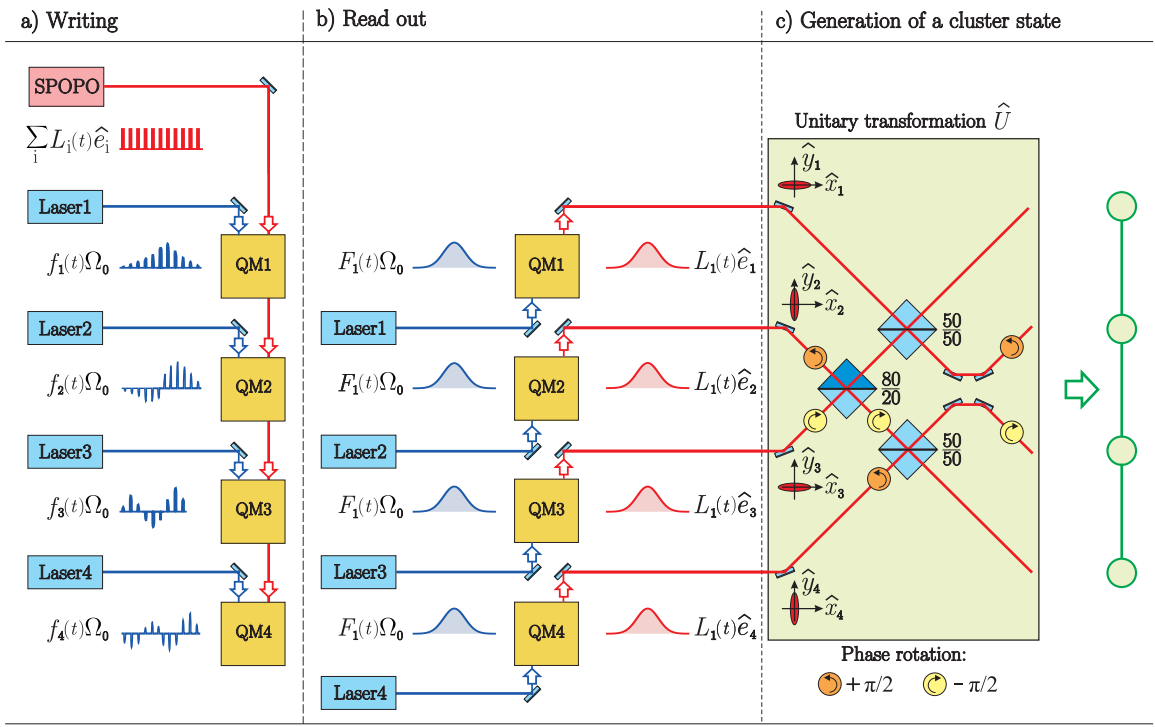}
\caption{Scheme of the thought experiment on generation of a light linear cluster consisting of four nodes.}
\label{cluster}
\end{center}
\end{figure}

Let us discuss the scheme of the thought experiment presented in Fig. \ref{cluster}.
The SPOPO signal pulse train enters the first memory cell $QM1$. Simultaneously, the input of the same cell is illuminated by a driving pulse train
from the \emph{laser 1}, matched to the signal field by pulse duration and the repetition period. The profile $F_1(t)$ of the driving is chosen such
a way (see Sec. \ref{iteration}) as to provide interaction of atoms with only the first supermode from the entire set of the SPOPO modes. As a
result, the first supermode scatters coherently on the atomic ensemble, mapping into spin wave distribution $B(z)$, and the rest of the signal field
passes through the first cell without interaction. A similar situation, but with the driving fields $F_i(t), i=2,3,4$, occurs on the subsequent
cells. Note that the profile of the spin wave formed in each of the cells appears to be the same, that allows us to effectively address the cell by
changing the shape of the driving (see Sec. \ref{kernels}).

During the readout driving fields with the same smooth profile $F_1(t)$ are entered the output faces (since the backward configuration is considered)
of each of the memory cells. Note that, unlike the writing, the reading field no longer has a pulse structure, so that the converted signal would
also have a smooth shape (coinciding with the first Hermite-Gaussian function in the considered example). Herewith, the light restored from each cell
would has the statistics of the radiation recorded on it and, consequently, this light would be squeezed in the $ \hat{y}$-quadrature for $k={1,3}$
and in the $\hat{x}$-quadrature for $k={2,4}$ \cite{Araujo, Roslund}.

Let us pay attention, since here we use quantum memory not for storage of a signal, but only for its conversion, the requirements to the storage time
of a memory cell can be minimal. At the same time, the fact that the cell is able to store a quantum state for some time, can be considered as an
additional advantage for performing logical operations on the cluster in the future.

Let us follow the constructing a joint multimode state based on the light to be read, and check that it would be a cluster state. Field
$\hat{A}_{out,k}(t)$ at the output of each of the memory cells can be decomposed over Hermite-Gaussian modes similar to the input signal of the SPOPO
(\ref{12}). Then, according to the procedure described above,
\BY
&&\hat{A}_{out,k}(t)=L_1(t)\hat{e}_k+vac=L_1(t)(\hat{x}_k+i\hat{y}_k)+vac,\qquad k=1,2,3,4,
\EY
where the profiles of all excited oscillators are determined by the first Hermite-Gaussian function $L_1(t)$,  and the index $k$  indicates the
number of the memory cell, where the light has been converted, and therefore the number of the supermode initially written into the cell. The
operator $\hat{e}_k$ coincides with the corresponding operator in decomposition (\ref{12}) and describes the quantum statistics of the stored mode.
Thus, the following conditions are satisfied: $\langle (\Delta \hat{y}_k)^2\rangle \rightarrow 0$ for $k=1,3$ and $\langle (\Delta
\hat{x}_k)^2\rangle \rightarrow 0$ for $k=2,4$. According to the experimental data \cite{Roslund},  the variances of the squeezed supermodes are
$\langle (\Delta \hat{y}_1)^2\rangle =0.10$, $\langle (\Delta \hat{x}_2)^2\rangle =0.12$, $\langle (\Delta \hat{y}_3)^2\rangle =0.14$, and $\langle
(\Delta \hat{x}_4)^2\rangle =0.18$, respectively.

The cluster that we want to construct (see Fig. \ref{cluster}) corresponds to a graph with an adjacency matrix
\BY
 V=\left(
     \begin{array}{cccc}
      0 &  1 & 0& 0 \\
     1 &0 & 1 & 0 \\
       0 & 1 & 0 &1 \\
       0 & 0 & 1 & 0 \\
     \end{array}
   \right).\L{V}
\EY
In compliance with \cite{Van Loock}, to create such a linear cluster with four light modes, squeezed alternately in different quadratures, one need
to perform over them the following unitary transformation $\vec{\hat{E}}(t)=U\vec{\hat{A}}_{out}(t)$, determined by the matrix $U$:
\BY
  U=\left(
     \begin{array}{cccc}
       \frac{1}{\sqrt{2}} &  \frac{i}{\sqrt{10}} & -\frac{2i}{\sqrt{10}}& 0 \\
      \frac{i}{\sqrt{2}} & \frac{1}{\sqrt{10}} & \frac{-2}{\sqrt{10}} & 0 \\
       0 & \frac{-2i}{\sqrt{10}} & \frac{-i}{\sqrt{10}} &\frac{1}{\sqrt{2}} \\
       0 & \frac{2}{\sqrt{10}} & \frac{1}{\sqrt{10}} & \frac{-i}{\sqrt{2}} \\
     \end{array}
   \right).\L{U}
\EY
This analytical expression can be considered as a step-by-step action of linear operators (see Appendix \ref{app1}), describing elements of the
optical scheme shown in Fig. \ref{cluster}, to the state of the field oscillators at the outputs of the memory cells.

Let us check whether the obtained multimode state actually is the declared linear cluster. According to the definition of the cluster state
\cite{Menicucci}, for the four-mode state one has to  specify four operators $\hat N_k, k=1...4$, constructed on the basis of canonical variables
$\{\hat p_k, \hat q_k\}$ as follows
\BY
&& \hat{N}_k= \hat{p}_k-\sum_{m=1}^4C_{km}\hat{q}_m,\L{nul}
\EY
and the action of which on the initial state (before the $U$ transformation) has to be zero. Due to this property such operators are called
\emph{nullifiers}. The coefficients $C_{km}$ can be different depending on the graph topology.

Since we solve the problem in the Heisenberg representation, we have to analyze variances of the nullifiers $\langle(\Delta\hat{N}_k)^2\rangle$, and
check whether these values vanish. Quadratures of the excited oscillators $\{\hat{\tilde x}_k, \hat{\tilde y}_k\}$, associated with the squares of
the fields at the outputs of the memory cells $\{\hat{x}_k,\hat{y}_k\}$ by the transformation (\ref {U}), have to be taken as canonical variables,
and the coefficients in the summation (\ref{nul}) are determined by the cluster graph, i.e. coincide with the elements of the matrix $V$ (\ref{V}).

As a result, we get:
\BY
&&\langle(\Delta\hat{N}_1)^2\rangle = \frac{2}{\sqrt{2}}\;\langle (\Delta \hat{y}_1)^2\rangle \rightarrow 0\\
&&\langle(\Delta\hat{N}_2)^2\rangle= \left(-\frac{5}{\sqrt{10}}\;\langle (\Delta \hat{y}_3)^2\rangle
-\frac{1}{\sqrt{2}}\;\langle (\Delta \hat{x}_4)^2\rangle\right) \rightarrow 0\\
&&\langle(\Delta\hat{N}_3)^2\rangle= \left(-\frac{5}{\sqrt{10}}\;\langle (\Delta \hat{x}_2)^2\rangle
+\frac{1}{\sqrt{2}}\;\langle (\Delta \hat{y}_1)^2\rangle\right) \rightarrow 0\\
&&\langle(\Delta\hat{N}_4)^2\rangle= -\frac{2}{\sqrt{2}}\;\langle (\Delta \hat{x}_4)^2\rangle  \rightarrow 0.
\EY

Thus, providing good squeezing in the initial SPOPO radiation, employing the mode separation and shaping with the help of the quantum memory cells,
we can construct a cluster state with chosen topology.

\section{Conclusion}

In this paper, we demonstrated how to convert the multimode quantum state of light, irradiated by a synchronously pumped optical parametric
oscillator, to create a cluster state. It should be noted that SPOPO radiation is a rather complex object to use: it is broadband, that makes
detection difficult, and has a complex comb structure, that is not easy to be matched by a homodyne. The proposed here procedure not only separates
the squeezed supermodes of the SPOPO, but also converts comb profiles of the modes into smooth and convenient for the further application ones. We
showed that, using different driving fields during the writing and readout and choosing their profiles so that in both cases the interaction turns
out to be with the same spin wave, we can convert the shape of the signal keeping its quantum statistics unchanged.

In the article, we discussed the case when the output field, read from each of the memory cells, has the same profile $L_1(t)$. Of course, as the
obtained calculations imply, we are not limited to this case only. We can convert one mode to another for the first six SPOPO supermodes. Moreover,
we can convert any linear combinations of these supermodes. This allows one to perform linear transformations over the selected modes directly during
the writing/readout process. An important condition here is addressing to the same spin wave.

The number of supermodes applicable to perform such manipulations is determined by the writing time $T_W$. It is well known that Hermite-Gaussian
modes constitute an orthonormal set on an infinite interval. However, conditions of the quantum memory work impose us to consider only a finite
intervals for the interaction times between fields and an atomic ensemble. That is why, choosing a certain value for $T_W$, we are limited to a set
of those functions that can be considered with sufficient accuracy orthogonal on the chosen interval. We performed calculations according to the
experimental parameters of the system presented in \cite{Roslund}.

Note once more that due to the fact that in our study the memory cell is used not for storing the quantum state of a signal, but for its noiseless
shaping, the requirements to the storage time are minimal, that simplifies experimental implementation. In particular, this places rather lenient
conditions on the atomic ensemble temperature, because thermal motion would not destroy the created distribution of the spin wave during the short
times. At the same time, having the resource of longer storage of the signal in the cell, we can use it in future when performing logical operations
with the cluster, implying that these operations require some time to be carried out.
\\

The reported study was supported by RFBR (Grants 15-02-03656a, 16-02-00180a).




\appendix
\section{Unitary transformation of fields for creating a cluster state. Clarification to the optical scheme}\L{app1}

The unitary transformation $U$, generating the cluster state, can be derived as a matrix decomposition:
\BY
U=F3\cdot F2 \cdot BS1 \cdot BS2 \cdot F3 \cdot F4 \cdot BS3\cdot F3\cdot F2,
\EY
where each element corresponds to an element of the optical scheme, implementing one of the linear transformations \cite{Van Loock}. First, according
to Fig. \ref{cluster},  beams $2$ and $3$ suffer the phase shifts by the angles of $\pi/2$ and $-\pi/2$, respectively. These operations correspond to
the matrixes $ F2 $ and $ F3 $:
\BY
F2=\left(
      \begin{array}{cccc}
        1 & 0 & 0 & 0 \\
        0 & i & 0 & 0 \\
        0 & 0 & 1 & 0 \\
        0 & 0 & 0 & 1 \\
      \end{array}
    \right), \qquad
F3=\left(
      \begin{array}{cccc}
        1 & 0 & 0  & 0 \\
        0 & 1 & 0  & 0 \\
        0 & 0 & -i & 0 \\
        0 & 0 & 0  & 1 \\
      \end{array}
    \right).
\EY
Then, these two beams are mixed on a beamsplitter with a transmission coefficient of $ 0.8 $. Such a mixing is described by the matrix $ BS3 $:
\BY
BS3=\left(
       \begin{array}{cccc}
         1 & 0 & 0 & 0 \\
         0 & \frac{1}{\sqrt{5}} & \frac{2}{\sqrt{5}} & 0 \\
         0 & \frac{2}{\sqrt{5}}& \frac{-1}{\sqrt{5}} & 0 \\
         0 & 0 & 0 & 1 \\
       \end{array}
     \right).
\EY
Beams $3$ and $4$ undergo the phase shifts by the angles of $-\pi/2$ and $\pi/2$, respectively. The shifting of the third beam is described by the
matrix $F3$, and we use matrix $F4$ to describe the shifting in the fourth channel:
\BY
F4=\left(
      \begin{array}{cccc}
        1 & 0 & 0 & 0 \\
        0 & 1 & 0 & 0 \\
        0 & 0 & 1 & 0 \\
        0 & 0 & 0 & i \\
      \end{array}
    \right).
\EY
Operators
\BY
BS1=\left(
       \begin{array}{cccc}
         \frac{1}{\sqrt{2}} & \frac{1}{\sqrt{2}} & 0 & 0\\
         \frac{1}{\sqrt{2}} & \frac{-1}{\sqrt{2}} & 0 & 0 \\
         0 & 0 & 1 & 0 \\
         0 & 0 & 0 & 1 \\
       \end{array}
     \right), \qquad
BS2=\left(
                \begin{array}{cccc}
                  1 & 0 & 0                  & 0 \\
                  0 & 1 & 0                  & 0 \\
                  0 & 0 & \frac{1}{\sqrt{2}} & \frac{1}{\sqrt{2}} \\
                  0 & 0 & \frac{1}{\sqrt{2}} & \frac{-1}{\sqrt{2}} \\
                \end{array}
              \right)
\EY
describe the mixing of two light waves (the first with the second and the third with the fourth, respectively) on symmetrical beamsplitters with a
transmission coefficient of $ 0.5 $.

Finally, beams $ 2 $ and $ 3 $  suffer again phase shifts by the angles of $\pi/2$ and $-\pi/2$, respectively, that is described by the matrixes $ F2
$ and $ F3 $.

As a result, the transformation $ U $ can by derived as follows:
\BY
 \left(
   \begin{array}{c}
 \hat{E}_1(t) \\
     \hat{E}_2(t) \\
      \hat{E}_3(t) \\
      \hat{E}_4(t) \\
   \end{array}
 \right)=U\;   \left(\begin{array}{c}
     \hat{A}_{out,1} (t) \\
    \hat{A}_{out,2} (t) \\
     \hat{A}_{out,3}(t)  \\
  \hat{A}_{out,4}(t) \\
   \end{array}\right) = L_1(t)\left(
           \begin{array}{c}
             \frac{1}{\sqrt{2}}\;\hat{x}_1- \frac{1}{\sqrt{10}}\;\hat{y}_2 +\frac{2}{\sqrt{10}}\;\hat{y}_3
             +i\left( \frac{1}{\sqrt{10}}\;\hat{x}_2-\frac{2}{\sqrt{10}}\;\hat{x}_3+ \frac{1}{\sqrt{2}}\;\hat{y}_1\right)\\
             \frac{1}{\sqrt{10}}\;\hat{x}_2- \frac{2}{\sqrt{10}}\;\hat{x}_3- \frac{1}{\sqrt{2}}\;\hat{y}_1
              +i\left(  \frac{1}{\sqrt{2}}\;\hat{x}_1+ \frac{1}{\sqrt{10}}\;\hat{y}_2- \frac{2}{\sqrt{10}}\;\hat{y}_3\right)\\
             \frac{1}{\sqrt{2}}\;\hat{x}_4+ \frac{2}{\sqrt{10}}\;\hat{y}_2+ \frac{1}{\sqrt{10}}\;\hat{y}_3
             +i\left( -\frac{2}{\sqrt{10}}\;\hat{x}_2- \frac{1}{\sqrt{10}}\;\hat{x}_3+ \frac{1}{\sqrt{2}}\;\hat{y}_4\right)\\
             \frac{2}{\sqrt{10}}\;\hat{x}_2+ \frac{1}{\sqrt{10}}\;\hat{x}_3+\frac{1}{\sqrt{2}}\;\hat{y}_4
              +i\left( -\frac{1}{\sqrt{2}}\;\hat{x}_4+ \frac{2}{\sqrt{10}}\;\hat{y}_2+\frac{1}{\sqrt{10}}\;\hat{y}_3\right)\\
           \end{array}
         \right)+vac.
 \EY

\end{document}